\RequirePackage{amsmath}
\documentclass[runningheads]{llncs}

\usepackage{tcolorbox}
\usepackage{etoolbox}

\newcounter{observationcounter}

\newif\ifEditMode

 \EditModetrue

\usepackage{preamble}
\usepackage{aliases}

\usepackage{pifont}

\begin{document}

\title{Fairness in Token Delegation: Mitigating Voting Power Concentration in DAOs}
\titlerunning{Fairness in Token Delegation}

\author{Johnnatan Messias\inst{1}
\and Ayae Ide\inst{2}}

\institute{{Max Planck Institute for Software Systems (MPI-SWS)}
\and {Pennsylvania State University}}

\authorrunning{J. Messias and A. Ide}

\maketitle
\begin{abstract}
  
Decentralized Autonomous Organizations (DAOs) aim to enable participatory governance, but in practice face challenges of voter apathy, concentration of voting power, and misaligned delegation. Existing delegation mechanisms often reinforce visibility biases, where a small set of highly ranked delegates accumulate disproportionate influence regardless of their alignment with the broader community. In this paper, we conduct an empirical study of delegation in DAO governance off-chain discussions from 14 DAO forums. We develop a methodology to link forum participants to on-chain addresses, extract governance interests using large language models, and compare these interests against delegates' historical behavior. Our analysis reveals that delegations are frequently misaligned with token holders' expressed priorities and that current ranking-based interfaces exacerbate power concentration. We argue that incorporating interest alignment into delegation processes could mitigate these imbalances and improve the representativeness of DAO decision-making.
To support future research, we will release our dataset and code in a public repository. 

\end{abstract}

\renewcommand{\point}[1]{\par\smallskip\noindent\textbf{#1.} }
\newcommand{\ayae}[1]{{\color{orange} { #1}}}

\section{Introduction}
\label{sec:intro}

\glspl{DAO} have become a key governance mechanism in the blockchain ecosystem, enabling token holders to participate in critical decisions that shape the direction of decentralized protocols. Here individuals or a collective of individuals can propose and vote on changes to the \glspl{DAPP} that run atop a blockchain. In contrast to traditional corporate governance structures, \glspl{DAO} aim to foster transparency, community ownership, and collective decision-making without relying on centralized intermediaries. However, despite these aspirations, real-world implementations of \gls{DAO} governance reveal significant shortcomings that threaten both their functionality and legitimacy, putting in check the decentralization motto of the blockchain ecosystem.

Recent studies~\cite{feichtinger2024sokattacksdaos,messias2024understandingblockchaingovernanceanalyzing,sharma2023unpacking} show that a large portion of voting power in many \glspl{DAO} is highly concentrated within a few delegates, resulting in skewed decision-making outcomes. This centralization often arises not from malicious manipulation but from structural and interface-level biases such as default sorting by voting power on delegation platforms like Tally (a popular platform currently in use by leading projects in the crypto ecosystem, including Arbitrum, Compound, Uniswap, \gls{ENS}, and ZKsync). At the same time, a majority of token holders abstain from voting altogether, contributing to voter apathy and weakening the diversity of voices in the governance process~\cite{feichtinger2024sokattacksdaos,messias2024understandingblockchaingovernanceanalyzing}. These issues not only undermine the promise of decentralization but also introduce systemic vulnerabilities, including governance capture, protocol ossification, and reduced adaptability to the best interests of the protocol's community.

\begin{figure}[tb]
\centering
\includegraphics[width=1\onecolgrid]{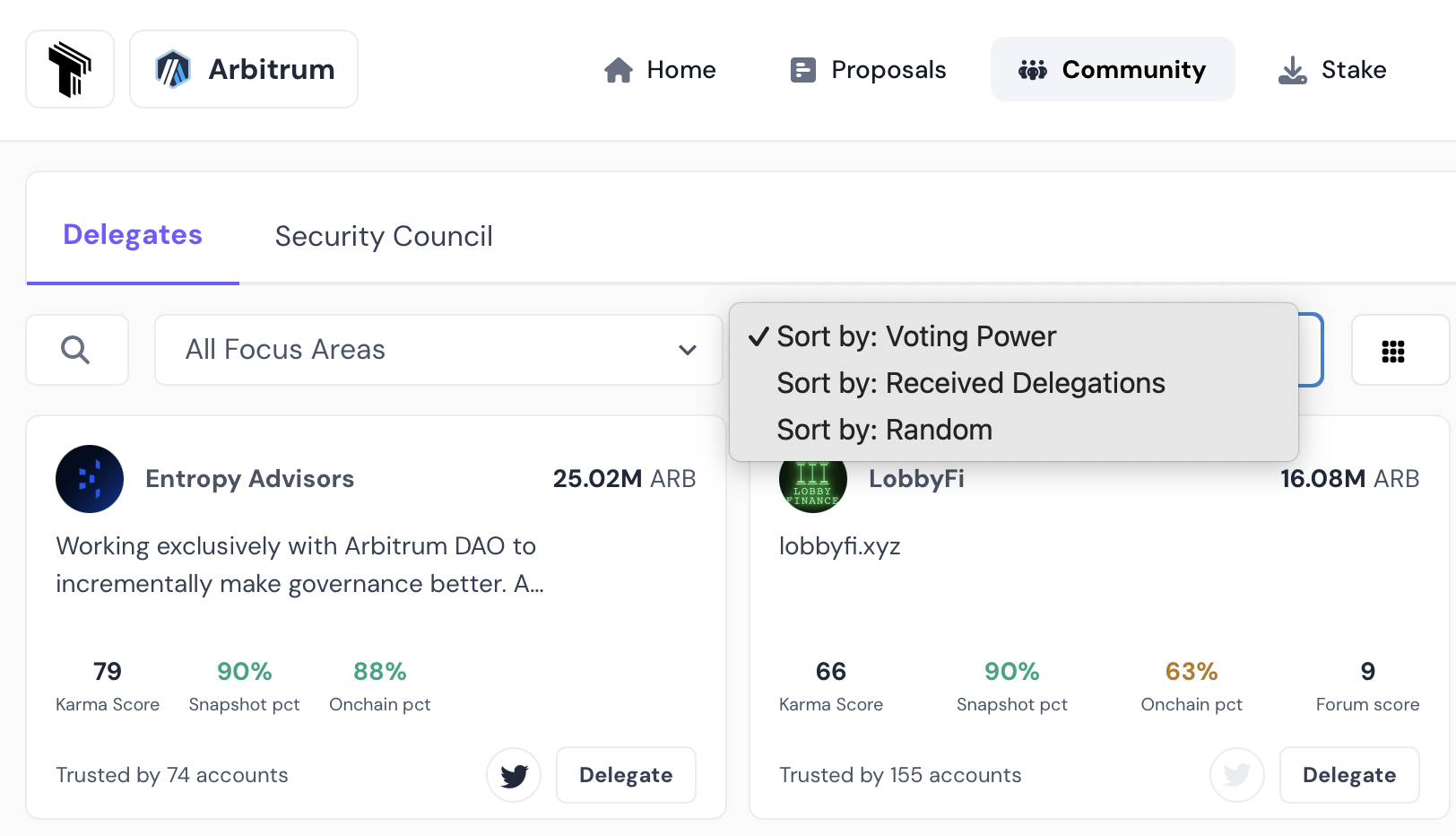}
\caption{ 
Delegation interface on Tally.
}
\label{fig:tally-platform}
\end{figure}

In order to mitigate the issue of voting participation, many \glspl{DAO} allow the users possessing tokens (or voting power) to delegate their tokens (or voting power) to another participant. This allows for something like a liquid or representative democracy where participants can vote on behalf of other users who do not wish to exercise their voting power~\cite{behrens2017origins,carroll1884principles,blum2016liquid}. This type of delegated voting was originally proposed as a way to mitigate participation fatigue in token-based governance by allowing token holders to assign their voting rights to more active participants. While this delegation mechanism increases turnout in theory, its current implementation often exacerbates inequality and misalignment. Delegates who accumulate early support or possess high visibility are more likely to attract additional votes, regardless of whether their views align with the broader community. This creates a self-reinforcing cycle where powerful delegates become even more influential, a dynamic reminiscent of the ``rich-get-richer'' phenomenon~\cite{Rich-gets-Richer@Wikimedia}. Furthermore, current interfaces like Tally (see Figure~\ref{fig:tally-platform}) offer little guidance on how to identify delegates that truly represent the token holder's values, preferences, or policy goals, while allowing their users to just rank-order blockchain delegates based on \stress{voting Power}, the amount of voting power the delegate has; \stress{received delegations}, the number of accounts that delegated their voting power to that delegate; and \stress{random}, supposedly ranking these accounts at random. However, prior research suggests that ranking order can influence user behavior, much like how product placements on Amazon drive revenue for certain sellers~\cite{Dash-Marketplaces,Dash-Alexa,Dash-FaiRIR}, raising concerns about whether delegations truly reflect users' values and fairness.

In this direction, this paper addresses these challenges by providing an
empirical study of \textit{interest-aligned delegation} in \gls{DAO} governance. 
Instead of examining delegation solely through the lens of popularity or 
delegate visibility, we investigate how token holders and delegates align 
in practice, drawing on shared values, historical voting patterns, and 
topic preferences extracted from off-chain governance forum discussions. 
Inspired by real-world advisory systems used in political elections 
(e.g., \textit{Wahl-O-Mat} in Germany~\cite{Bundeszentrale@Wahl}), our analysis 
seeks to evaluate to what extent delegations reflect underlying interests 
and how alignment (or misalignment) impacts representational quality 
and the concentration of voting power.

Our study relies on a multi-modal analysis of both on-chain and off-chain 
data. We leverage full Ethereum archival data to reconstruct governance 
actions such as proposal creation, vote casting, and delegation changes. 
Simultaneously, we process large volumes of off-chain posts using 
\glspl{LLM} techniques to extract topics, sentiment, and ideological signals. 
By combining these sources, we build behavioral profiles of both delegates 
and token holders, enabling a principled measurement of alignment in 
delegation choices.
Our work makes the following key contributions:

\StarCase We build a comprehensive dataset of off-chain governance discussions from platforms like Tally and official forums discussion, 
focusing on 14 \glspl{DAO}.

\StarCase We propose a methodology to quantify interest alignment between 
token holders and potential delegates by integrating historical off-chain discussion signals.

\StarCase We plan to share our code and dataset in 
the final version of our paper in a public repository to enable the 
reproducibility of our results.

Our work can help mitigate delegation imbalances and reduce 
voting power concentration in \glspl{DAO}. Our findings also have practical 
relevance, as they could inform leading platforms such as Tally in designing

\section{Related Work}
\label{sec:related_work}

\glspl{DAO} are a central focus of research on blockchain governance. Over the years, several studies have investigated the strengths and weaknesses of \gls{DAO} governance structures, particularly their decentralization and the impact of delegation mechanisms.

\paraib{Governance Design and Trustworthy \glspl{DAO}}
The governance of DAOs is designed to empower token holders to make decisions about the organization's operations, with a focus on transparency and decentralization.
Okutan et al.~\cite{Okutan2025DemocracyDAOs} in their recent empirical work analyzes governance and participation patterns in 14 Internet Computer SNS DAOs, finding high engagement but also highlighting design elements that influence decentralization outcomes. 

\paraib{Decentralization Metrics}
The challenge of achieving true decentralization in \gls{DAO} voting mechanisms is significant. For instance, Messias et al.~\cite{messias2024understandingblockchaingovernanceanalyzing} show that there is often a high concentration of voting power in \glspl{DAO}, which leads to concerns about the equitable distribution of governance tokens. This study on Compound and Uniswap exposes the centralization risks due to token holders' ability to propose and approve changes to smart contracts. Small token holders may face significant barriers, both in terms of costs and influence, thereby undermining the decentralization ideals of \glspl{DAO}. Austgen et al.~\cite{austgen2023daodecentralizationvotingblocentropy} propose \gls{VBE} as a metric to quantify the decentralization effects of voting blocs, highlighting risks of systemic bribery and vote-buying. 
Complementary evidence comes from Weidener et al.~\cite{Weidener2025DelegatedVotingReview}, who synthesize DAO delegated voting implementations and metrics across projects, noting that without constraints, delegation can exacerbate centralization despite increasing turnout.

\paraib{DAO Vulnerabilities and Security}
The security of \glspl{DAO}, particularly in the context of governance attacks, has also been a focus of recent research. Feichtinger et al.~\cite{feichtinger2024sokattacksdaos} identify various types of attacks that have targeted \glspl{DAO}, including bribery, token control, and protocol vulnerabilities. These attacks often exploit the inherent flaws in \gls{DAO} governance mechanisms, especially when voting power is concentrated in the hands of a few entities. Other studies focus on the risks of manipulation in delegation mechanisms. For example, Alouf-Heffetz et al.~\cite{Alouf-Heffetz2024ControllingDelegations} model the complexity of controlling delegation graphs to influence outcomes in liquid democracy, while Fang et al.~\cite{Fang2024DelegationICIS} empirically show that being delegated increases engagement but can also lead delegatees to follow the majority, undermining decentralization.

\paraib{Delegated Voting and Liquid Democracy}
Delegated voting (or \emph{liquid democracy}) offers a flexible alternative to direct voting, aiming to combine broad participation with scalable decision-making~\cite{blum2016liquid}. However, multiple works point to trade-offs in fairness, accountability, and vulnerability to concentration. Weidener et al.~\cite{Weidener2025DelegatedVotingReview} identify design choices across DAOs that exacerbate or mitigate these effects. Nazirkhanova et al.~\cite{Nazirkhanova2025Kite} propose \emph{Kite}, a privacy-preserving delegation protocol based on \glspl{ZKP}, allowing public or private delegation without exposing delegator identities. Shah et al.~\cite{Shah2025ArtificialDelegates} introduce artificial delegates to stand in for absent voters, improving fairness metrics such as quota compliance and reducing Gini influence in perpetual voting. Alouf-Heffetz et al.~\cite{Alouf-Heffetz2024CostPerspectiveLD} study liquid democracy from a cost-minimization perspective, showing how budget constraints and delegation chain limits can maintain representation while preventing concentration.

\paraib{Empirical Analyses of Delegation}
Several recent empirical works examine how delegation unfolds in practice. Bongaerts et al.~\cite{bongaerts2025vote} analyze Uniswap governance, finding that entities with minimal self-owned voting power and those affiliated with major venture firms attract disproportionate delegations, and that prior proposal success correlates with delegation inflows. Messias et al.~\cite{messias2024understandingblockchaingovernanceanalyzing} find that delegation patterns often concentrate influence among a small set of actors, shaping proposal outcomes and raising concerns about the representativeness of decentralized governance.
Fang et al.~\cite{Fang2024DelegationICIS} highlight the double-edged nature of delegation, where engagement gains are offset by tendencies to conform to the majority. \stress{These observations align with our own findings that naive delegate discovery mechanisms can amplify existing power asymmetries.}

\paraib{Summary of Our Work}
While prior studies have documented the risks of concentration and the  trade-offs of delegation, they leave open questions about how well  delegations reflect the underlying preferences of token holders. Our work addresses this gap through an empirical study of \gls{DAO} delegation that leverages off-chain governance discussions across 14 DAOs. By applying \glspl{LLM} to forum data, we extract voter interests and compare them against delegate behavior, providing the first large scale measurement of interest alignment in DAO delegation. We argue that making alignment more transparent could help redistribute delegations more equitably and improve the fairness of DAO decision-making.

\section{Background and Data Collection}
\label{sec:methodology}

\begin{figure}[tb]
\centering
\includegraphics[width=1\onecolgrid]{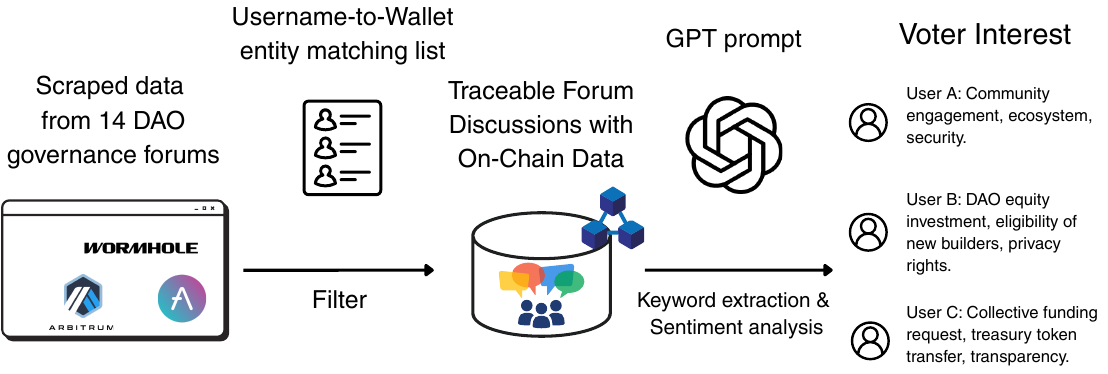}
\caption{
Overview of our interest-aligned delegation pipeline. Forum data from 14 DAOs are linked to on-chain addresses via \stress{username-to-wallet mappings}. We extract linguistic and sentiment features and use \gls{GPT}-based prompts to derive voter interest categories for delegate recommendations.
}
\label{fig:gpt-pipeline}
\end{figure}

In this section, we provide background on DAO token ownership and delegation, and describe how we collect and process our dataset. 
Our analysis uses \emph{off-chain} data, obtained from 14 official governance forums. 
Figure~\ref{fig:gpt-pipeline} summarizes our pipeline. 
We first collect forum discussions and map their username-to-wallet that could be used with on-chain governance voters actions. 
Only reliably traceable accounts are retained. 
Next, we apply \gls{NLP}  methods, keyword extraction and sentiment analysis, to capture issue-specific signals from user posts. 
These signals are then translated, via \gls{GPT}-based prompting, into interpretable categories of voter interests such as ecosystem security or treasury transparency. 
The resulting profiles provide a structured view of token holders' governance preferences and serve as the foundation for designing fairness-aware delegation mechanisms.

\subsection{Token Ownership and Delegation Roles}

DAO governance typically involves three key concepts: \emph{token holders}, \emph{delegators}, and \emph{delegates}. 
Understanding the distinctions among these roles is key for interpreting the concentration patterns.

\paraib{Token holders} 
Any address that owns governance tokens of a protocol is considered a token holder. 
Ownership provides the right to vote directly on proposals or to delegate voting power to another participant. 
Token distributions are often highly unequal, where early investors, foundations, or large funds control disproportionately large fractions of the supply.

\paraib{Delegators} 
A delegator is a token holder who chooses not to vote directly but instead transfers their voting power to another account. 
Delegators can assign all of their voting power to a single delegate or, in some protocols, split it across multiple delegates. 
In practice, most delegators assign their entire balance to one delegate, which simplifies participation but concentrates influence. 

\paraib{Delegates} 
Delegates are the recipients of delegated voting power. 
They act as representatives on behalf of one or more token holders and are responsible for casting votes in governance proposals. 
Delegates may or may not hold a significant number of tokens themselves; their influence often comes from the aggregated power delegated by others. 
This makes them highly visible actors in DAO governance, frequently appearing at the top of delegation interfaces such as Tally.

\subsection{Forum Data Acquisition and Identify Resolution}

\begin{table*}[t]
\centering
\small
\caption{Overview of Selected DAO Protocols obtained from CoinMarketCap in March 2026~\cite{DAOs-CoinMarketCap}.}
\label{tab:dao_overview}
\resizebox{\textwidth}{!}{%
\begin{tabular}{lcccccccc}
\toprule
\multirow{2}{*}{\thead{Protocol}} & \multicolumn{1}{c}{\thead{Launch}} & \multicolumn{1}{c}{\thead{Governance}} & \thead{Focus Area} &  \multicolumn{1}{c}{\thead{Market}} & \thead{Blockchain}  & \multicolumn{1}{c}{\thead{Token}} & \multicolumn{1}{c}{\thead{Governance}}\\
& \thead{Year} & \thead{Token} & & \thead{Cap} & &  \thead{Contract} & \thead{Contract}\\
\midrule
Aave~\cite{frangella2022aave} & 2017 & AAVE & Lending and Borrowing & \$1.61B & Ethereum &  \href{https://etherscan.io/address/0x7fc66500c84a76ad7e9c93437bfc5ac33e2ddae9}{0x7fc6$\cdots$dae9} & \href{https://etherscan.io/address/0xec568fffba86c094cf06b22134b23074dfe2252c}{0xec56$\cdots$252c} \\
Compound~\cite{leshner2019compound} & 2018 & COMP & Lending and Interest Rates & \$186.29M & Ethereum &  \href{https://etherscan.io/address/0xc00e94cb662c3520282e6f5717214004a7f26888}{0xc00e$\cdots$6888} & \href{https://etherscan.io/address/0xc0da02939e1441f497fd74f78ce7decb17b66529}{0xc0da$\cdots$6529} \\
ENS~\cite{ENS} & 2017 & ENS & Decentralized Domain Names & \$220.79M & Ethereum & \href{https://etherscan.io/address/0xc18360217d8f7ab5e7c516566761ea12ce7f9d72}{0xc183$\cdots$9d72} & \href{https://etherscan.io/address/0x323a76393544d5ecca80cd6ef2a560c6a395b7e3}{0x323a$\cdots$b7e3} \\
Nouns~\cite{Nouns} & 2021 & Nouns & NFT Auction & --- & Ethereum & \href{https://etherscan.io/address/0x9c8ff314c9bc7f6e59a9d9225fb22946427edc03}{0x9c8f$\cdots$dc03} & \href{https://etherscan.io/address/0x6f3e6272a167e8accb32072d08e0957f9c79223d}{0x6f3e$\cdots$223d} \\
Uniswap~\cite{adams2021uniswap} & 2018 & UNI & Decentralized Exchange (DEX) & \$2.19B & Ethereum &  \href{https://etherscan.io/address/0x1f9840a85d5af5bf1d1762f925bdaddc4201f984}{0x1f98$\cdots$f984} & \href{https://etherscan.io/address/0x408ed6354d4973f66138c91495f2f2fcbd8724c3}{0x408e$\cdots$24c3} \\
\bottomrule
\end{tabular}}
\end{table*}

\begin{table*}[t]
\centering
\small
\caption{Summary of events related to the protocols' tokens and their DAO contracts that we gathered from the blockchain from their inception to Jun. 19, 2025 (block \#\num{22740000}).}
\label{tab:events_tokens_governance}
\resizebox{\textwidth}{!}{%
\begin{tabular}{crrrrrrp{6.8cm}}
\toprule
\multicolumn{1}{c}{\thead{Contract}} & \multicolumn{1}{r}{\thead{Event}} & \multicolumn{5}{c}{\thead{\# of events}} & \multirow{2}{*}{\thead{Description}} \\
   \thead{Type}  &  \thead{Name}  & \thead{Uniswap}     & \thead{AAVE} & \thead{Nouns} & \thead{ENS}  & \thead{Compound}    &     \\
\midrule
\multirow{4}{*}{\thead{Token}} & \stress{Approval} & \num{933881} & \num{1237982} & \num{5191} & \num{426955} & \num{281483}  &  \small{Standard ERC-20 approval event.} \\
 & \stress{DelegateChanged} & \num{56695} & \num{10580} & \num{1658} & \num{119781} & \num{17938} & \small{Emitted when an account changes its delegate.} \\
 & \stress{DelegateVotesChanged} & \num{174636} & --- & \num{13249} & \num{301665} & \num{87091} & \small{Emitted when a delegate account's vote balance changes.} \\
 & \stress{Transfer} & \num{5629750} & \num{3698930} & \num{8064} & \num{1227174} & \num{2377436} &  \small{Emitted when users/holders transfer their tokens to another address.} \\
\hline
\multirow{5}{*}{\thead{Governance}} & \stress{ProposalCanceled} & \num{24} & \num{41} & \num{116} & --- & \num{40}   &  \small{Emitted when a proposal is canceled.} \\
 & \stress{ProposalCreated} & \num{87} & \num{743} & \num{794} & \num{42} & \num{393} &  \small{Emitted when a new proposal is created.} \\
 & \stress{ProposalExecuted} & \num{59} & \num{670} & \num{437} & \num{38} & \num{317} & \small{Emitted when a proposal is executed in the TimeLock.} \\
 & \stress{ProposalQueued} & \num{59} & \num{682} & \num{446} & \num{39} & \num{322} &  \small{Emitted when a proposal is added to the queue in the TimeLock.} \\
 & \stress{VoteCast} & \num{55776} & --- & \num{24877} & \num{6644} & \num{16812} & \small{Emitted when a vote is cast on a proposal: \num{0} for against, \num{1} for in-favor, and \num{2} for abstain.} \\
\bottomrule
\end{tabular}}
\end{table*}

\paraib{On-chain Data}
We gathered governance events from five major protocols (Aave, Compound, ENS, Nouns, and Uniswap) covering their full history up to June 19, 2025. These protocols were selected based on market capitalization and availability of both delegation and governance forum data. To ensure complete coverage, we deployed Ethereum and Arbitrum archive nodes on a dedicated server (64 CPU cores, 252 GB RAM, 21 TB storage) and synchronized them with the respective blockchains. We used the \texttt{Web3.py} library to extract all governance-related transactions, including token transfers, delegate changes, proposal lifecycle events, and vote casting. Table~\ref{tab:dao_overview} summarizes the selected protocols, and Table~\ref{tab:events_tokens_governance} details the number of governance and token-related events.

\begin{table*}[t]
\centering
\caption{Summary of the data attributes gathered from each platform.}
\label{tab:data_attributes}
\resizebox{\textwidth}{!}{%
\begin{tabular}{lp{8cm}p{8cm}}
\toprule
\thead{Data Source} & \thead{Covered DAOs} & \thead{Collected Attributes} \\
\midrule
\textit{Governance forums} & 
Aave, ApeCoin, Arbitrum, Compound, \gls{ENS}, Jito, LidoDAO, Maple Finance, Nervos Network, UMA, Uniswap, Wormhole, yearn.finance, 0xProtocol &

\StarCase \textbf{Proposal metadata}: ID, title, creation date, author, and HTML-rendered text.

\StarCase \textbf{User metadata}: User ID, username, optional name for proposal/reply authors.

\StarCase  \textbf{Discussion posts}: All HTML-rendered replies, including message content, quoted text, and links. \\
\midrule
\textit{Tally} & 
Aave, Arbitrum, Compound, \gls{ENS}, Uniswap, Wormhole &

\StarCase \textbf{Delegates information}: User ID, username, X/Twitter handle, wallet address, voting power, and bio details. \\

\bottomrule
\end{tabular}}
\end{table*}

\begin{table*}[t]
\centering
\caption{Summary of the posts we gathered from DAO forums to Jul. 16, 2025.}
\resizebox{\textwidth}{!}{%
\begin{tabular}{lrrrrrrr}
\toprule
\multirow{2}{*}{\thead{Protocol}} & \multirow{2}{*}{\thead{\# of Proposals}} & \multirow{2}{*}{\thead{\# of Posts}} & \multirow{2}{*}{\thead{\# of Unique Users}} & \multicolumn{4}{c}{\thead{Posts per User}}\\
&&&& \thead{Min.} & \thead{Avg.} & \thead{Median} & \thead{Max.} \\
\midrule
Aave & 2692 & \num{22314} & 3222 & 1 & 6.93 & 1 & 1401\\
ENS & 2196 & \num{20395} & 2606 & 1 & 7.83 & 1 & 1773 \\
Arbitrum & 2154 & \num{29425} & 3614 & 1 & 8.14 & 1 & 750\\
NervosNetwork & 2068 & \num{10606} & 1087 & 1 & 9.76 & 2 & 443\\
ApeCoin & 1845 & \num{35888} & 1539 & 1 & 23.32 & 4 & 2096\\
Compound & 1300 & 8842 & 1055 & 1 & 8.38 & 2 & 1019\\
yearnfinance & 1270 & \num{13747} & 1896 & 1 & 7.25 & 2 & 528\\
LidoDAO & 1076 & 9217 & 1535 & 1 & 6.00 & 2 & 437\\
Uniswap & 894 & \num{11393} & 2215 & 1 & 5.14 & 1 & 398\\
UMA & 447 & 1252 & 254 & 1 & 4.93 & 1 & 333\\
0xProtocol & 160 & 1665	& 744 & 1 & 2.24 & 1 & 141\\
Wormhole & 99 & 211 & 99 & 1 & 2.13 & 1 & 13\\
Jito & 81 & 351 & 98 & 1 & 3.58 & 1 & 34\\
MapleFinance & 42 & 359 & 126 & 1 & 2.85 & 2 & 29\\
\bottomrule
\end{tabular}}
\label{tab:forum-dataset-summary}
\end{table*}

\paraib{Off-chain Forum Data}
We collected discussions from the governance forums of 14 \glspl{DAO}: Aave, ApeCoin, Arbitrum, Compound, ENS, Jito, LidoDAO, Maple Finance, Nervos Network, UMA, Uniswap, Wormhole, yearn.finance, and 0xProtocol. These \glspl{DAO} were chosen from the top 30 market capitalization (as of July 2025). For each forum, we collected all proposals and their associated discussions threads, including \stress{Proposal metadata} (ID, title, creation date, author, and rendered text), \stress{User metadata} (ID, username, and optional display name), and \stress{Discussion posts} (complete HTML-rendered replies, including quoted text and embedded links). We also queried their delegation information from Tally, including wallet addresses, voting power, X/Twitter social media handles, and bibliographies. We provide details in Table~\ref{tab:data_attributes} and Table~\ref{tab:forum-dataset-summary}.

Table~\ref{tab:forum-dataset-summary} reports statistics from 14 \gls{DAO}
governance forums, including proposals, posts, and unique users, as
well as per-user activity distributions. The dataset covers lending
protocols (Aave, Compound, Maple Finance), DeFi primitives
(Uniswap, yearn.finance), middleware (Arbitrum, Wormhole, Nervos),
and community projects (ENS, ApeCoin, LidoDAO, Jito, UMA,
0xProtocol).

Participation is highly skewed: most users post only once, while a
few dominate discussions (e.g., over \num{2000} posts in ApeCoin and
over \num{1700} in ENS). This mirrors on-chain governance patterns where a
small set of actors concentrates influence.

\paraib{Identity Resolution} To link off-chain forum users with on-chain accounts, we mapped usernames from governance forums to Tally delegate profiles and \gls{ENS} registration. This process relied on \gls{ENS} registry events, obtained by parsing contract \glspl{ABI} via Etherscan, to associate wallet addresses with \gls{ENS} names. Out of \num{86445} Tally-listed addresses, we successfully inferred \gls{ENS} domains for \num{72186} (83.5\%), providing cross-platform linkage between discussion activity and voting behavior.

\subsection{Data Labeling}

\paraib{Username-to-Wallet Mapping via Tally}
To associate discussions from governance forums with on-chain activities, we performed user-to-wallet mapping using public profile data available on Tally. Based on a set of matching heuristics described below, with specific examples provided in Table \ref{tab:labeling_examples}, we linked the identifiers obtained from Tally to those from governance forums, enabling us to associate off-chain forum users with on-chain wallet addresses. 
Table~\ref{tab:forum-matched-summary} summarizes the subset of forum data from Table~\ref{tab:forum-dataset-summary}  that we successfully \stress{entity-matched} with on-chain addresses from Tally. Despite partial matching rates, the resulting dataset remains broadly representative of the original forum activity.

\begin{table*}[t]
\centering
\caption{Examples of Forum–to-Tally User Entity Matching}
\label{tab:labeling_examples}
\resizebox{\textwidth}{!}{
\begin{tabular}{c | lllll | p{3cm}} 
\toprule
\thead{Label} & \thead{Username (forum)} & \thead{Optional name (forum)} & \thead{Mapped \texttt{.eth} name (Tally)} & \thead{Username (Tally)} &\thead{Twitter handle (Tally)} & \thead{Matching reason} \\
\midrule
\multirow{2}{*} \textit{High-confidence} & \textbf{adalovelace.eth} &  & \textbf{adalovelace.eth} & adalovelace & adalovelace & \multirow{2}{=}{Exact ENS match}\\
 & vitalikbuterin & \textbf{vitalik.eth} & \textbf{vitalik.eth} & vitalik & & \\
 
\midrule
\multirow{2}{*} \textit{Middle-confidence} & \textbf{adalovelace} &  & \underline{0xlovelace.eth} & \underline{0xada.eth} & \textbf{adalovelace123} & \multirow{2}{=}{Username and ENS alignment}\\
 & vitalik & \textbf{vitalikcrypto} & \underline{vitalikcrypto.eth} & \underline{vitalikcrypto.eth} & \textbf{vitalikcrypto} & \\
\midrule

\multirow{2}{*} \textit{Low-confidence} & viiitalik.eth & \textbf{vitalik} & 0xvitalik.eth & \textbf{vitalik}  & vitalikbuterin & \multirow{2}{=}{Naive name-based match}\\
 & viiitalik.eth & \textbf{vitalik} & N/A & \textbf{vitalik} & viiitalik2013 & \\

\midrule
 \multirow{2}{*} \textit{Manually verified} & \textbf{sushi} & sushi & N/A & \textbf{sushi} & & Profile icons match\\
 & \textbf{blockchainedu} & & blockchainedu.eth & ben & \textbf{blockchainedu} & Profile info match\\
 
\bottomrule
\end{tabular}}
\end{table*}

\begin{table*}[t]
\centering
\caption{Summary of posts from matched data between governance forums and Tally. (\%) indicates the match rate.}
\resizebox{\textwidth}{!}{%
\begin{tabular}{lrrrrrrrr}
\toprule
\multirow{2}{*}{\thead{Protocol}} &
\multirow{2}{*}{\thead{\# of Matched Proposals}} &
\multirow{2}{*}{\thead{\# of Matched Posts}} &
\multirow{2}{*}{\thead{\# of Matched\newline Unique Users}} &
\multicolumn{4}{c}{\thead{Posts per User}}\\
&&&& \thead{Min.} & \thead{Avg.} & \thead{Median} & \thead{Max.} \\
\midrule
Aave  & 653 (24.3\%) & 1659 (7.4\%) & 25 (0.8\%) & 1 & 66.36 & 3 & 981 \\
ENS  & 1582 (72.0\%) & 7040 (34.5\%) & 236 (9.1\%) & 1 & 29.83 & 2 & 593 \\
Arbitrum  & 880 (40.9\%) & 4290 (14.6\%) & 94 (2.6\%) & 1 & 45.64 & 3 & 541 \\
NervosNetwork  & 57 (2.8\%) & 91 (0.9\%) & 2 (0.2\%) & 1 & 45.50 & 45 & 90 \\
ApeCoin  & 108 (5.9\%) & 267 (0.7\%) & 18 (1.2\%) & 1 & 14.83 & 5 & 80 \\
Compound & 299 (23.0\%) & 774 (8.8\%) & 20 (1.9\%) & 1 & 38.70 & 10 & 250 \\
yearnfinance & 89 (7.0\%) & 182 (1.3\%) & 11 (0.6\%) & 1 & 16.55 & 2 & 108 \\
LidoDAO  & 57 (5.3\%) & 108 (1.2\%) & 21 (1.4\%) & 1 & 5.14 & 2 & 43 \\
Uniswap  & 196 (21.9\%) & 730 (6.4\%) & 38 (1.7\%) & 1 & 19.21 & 2 & 131 \\
UMA  & 0 (--\%) & 0 (--\%) & 0 (--\%) & -- & -- & -- & -- \\
0xProtocol  & 6 (3.8\%) & 24 (1.4\%) & 9 (1.2\%) & 1 & 2.67 & 1 & 14 \\
Wormhole & 33 (33.3\%) & 53 (25.1\%) & 18 (18.2\%) & 1 & 2.94 & 1 & 13 \\
Jito & 20 (24.7\%) & 40 (11.4\%) & 5 (5.1\%) & 1 & 8.00 & 8 & 17 \\
MapleFinance & 1 (2.4\%) & 1 (0.3\%) & 1 (0.8\%) & 1 & 1.00 & 1 & 1 \\
\bottomrule
\end{tabular}}
\label{tab:forum-matched-summary}
\end{table*}

\textcircled{\footnotesize{1}} \parab{\textit{High-confidence} (Exact ENS match)} This category includes cases where the forum username or optional name attributes exactly matches the \texttt{.eth} name derived from a Tally-linked wallet address. The ENS name must be uniquely associated with the on-chain address to qualify as high-confidence. We identified 284 such matches, accounting for 0.33\% of the \num{86445} delegates on Tally.

\textcircled{\footnotesize{2}} \parab{\textit{Middle-confidence} (Username and ENS alignment)} In these cases, the forum username or optional name exactly corresponds to the Tally username or Twitter handle displayed on Tally. Furthermore, if the matched Tally account username corresponds to a \texttt{.eth} name that can be derived from its wallet address, we treat the identity as moderately reliable. This yielded 88 matches (0.10\% of all delegates). 75 of these cases do not include \texttt{.eth} in the forum name but contain the prefix portion of the \texttt{.eth} name in the username, and the remaining 13 cases have forum names of at least eight characters, which we verified to be sufficiently distinctive.

\textcircled{\footnotesize{3}} \parab{\textit{Low-confidence} (Naive name-based match)} These matches are based on simple name matches between forum identifiers (e.g., username or optional name) and Tally identifiers (e.g., username or Twitter handle), without supporting evidence from ENS names. Multiple candidate matches may exist, and the ambiguity has not yet been resolved.  We identified \num{2067} such cases (2.39\%), but excluded low-confidence matches from our analysis due to their high ambiguity and the absence of supporting signals, which render the inferred connections unreliable.

\textcircled{\footnotesize{4}} \parab{\textit{Manually Verified}} We found 73 matches (0.08\%), which were confirmed through manual cross-referencing of multiple signals, including partial alignment with \texttt{.eth} name (e.g., forum-username.eth), profile icons, social media links, personal websites, and stated affiliations between the forum and Tally. In some cases, identities were further validated by user activity observed across both platforms.

\section{On-Chain Data Analysis of DAO Governance}
\label{sec:on-chain}

\begin{figure*}[t]
    \centering
    \begin{subfigure}{1\twocolgrid}
        \centering
        \includegraphics[width=\twocolgrid]{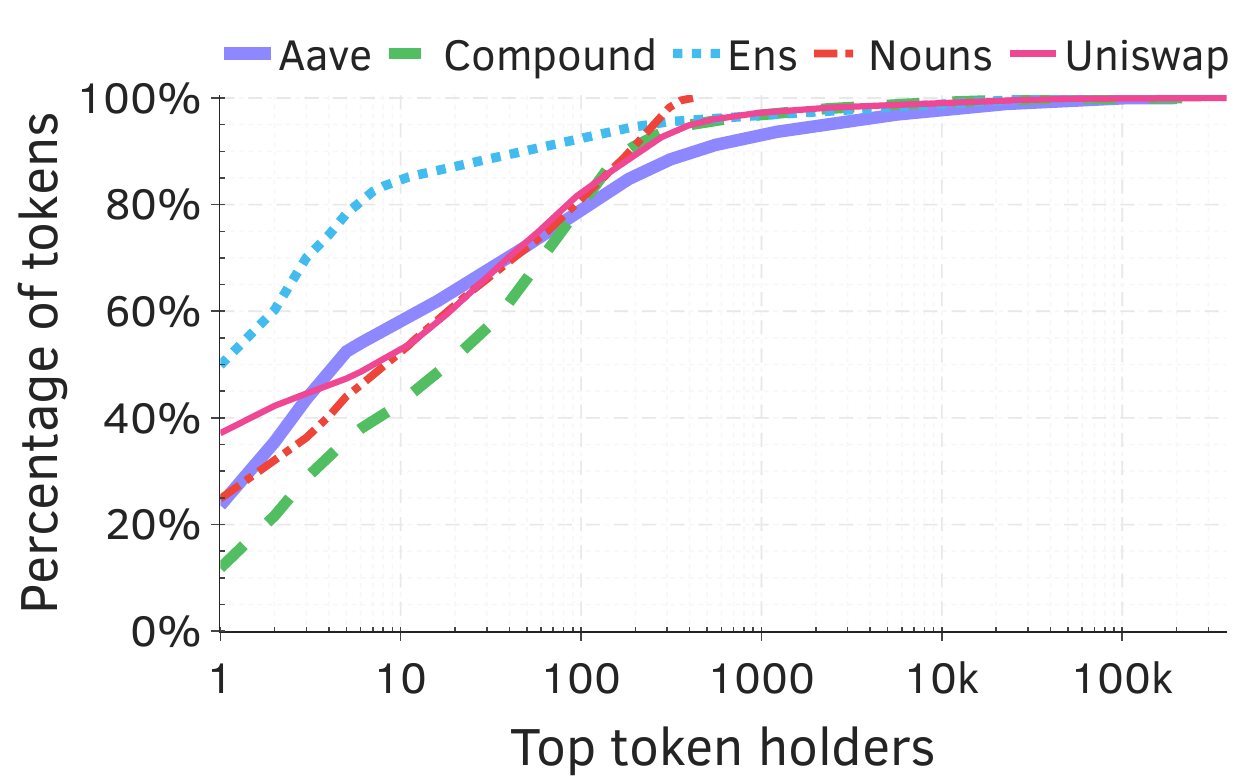}
        \caption{Top holders.}
        \label{fig:top-accounts-holders}
    \end{subfigure}
    \begin{subfigure}{1\twocolgrid}
        \centering
        \includegraphics[width=\twocolgrid]{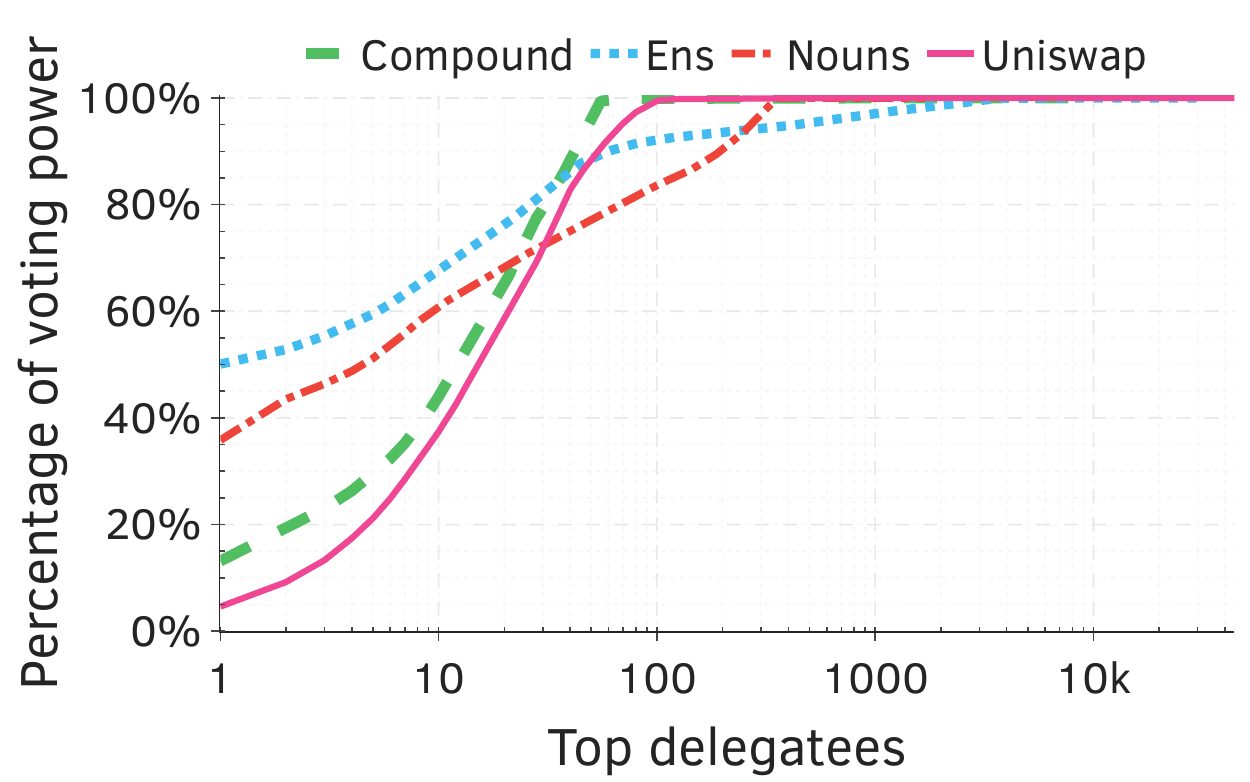}
        \caption{Top delegatees.}
        \label{fig:top-accounts-delegates}
    \end{subfigure}
    \begin{subfigure}{1\twocolgrid}
        \centering
        \includegraphics[width=\twocolgrid]{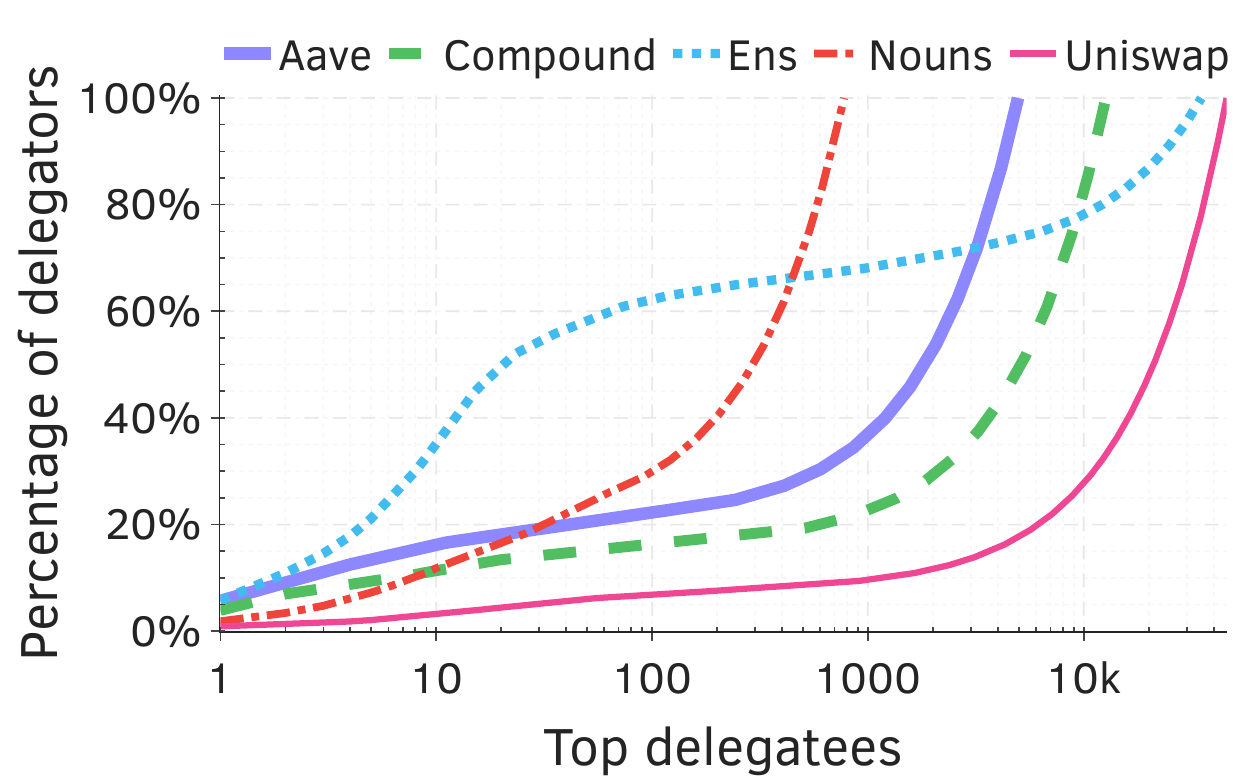}
        \caption{Top delegators.}
        \label{fig:top-accounts-delegators}
    \end{subfigure}
    \begin{subfigure}{1\twocolgrid}
        \centering
        \includegraphics[width=\twocolgrid]{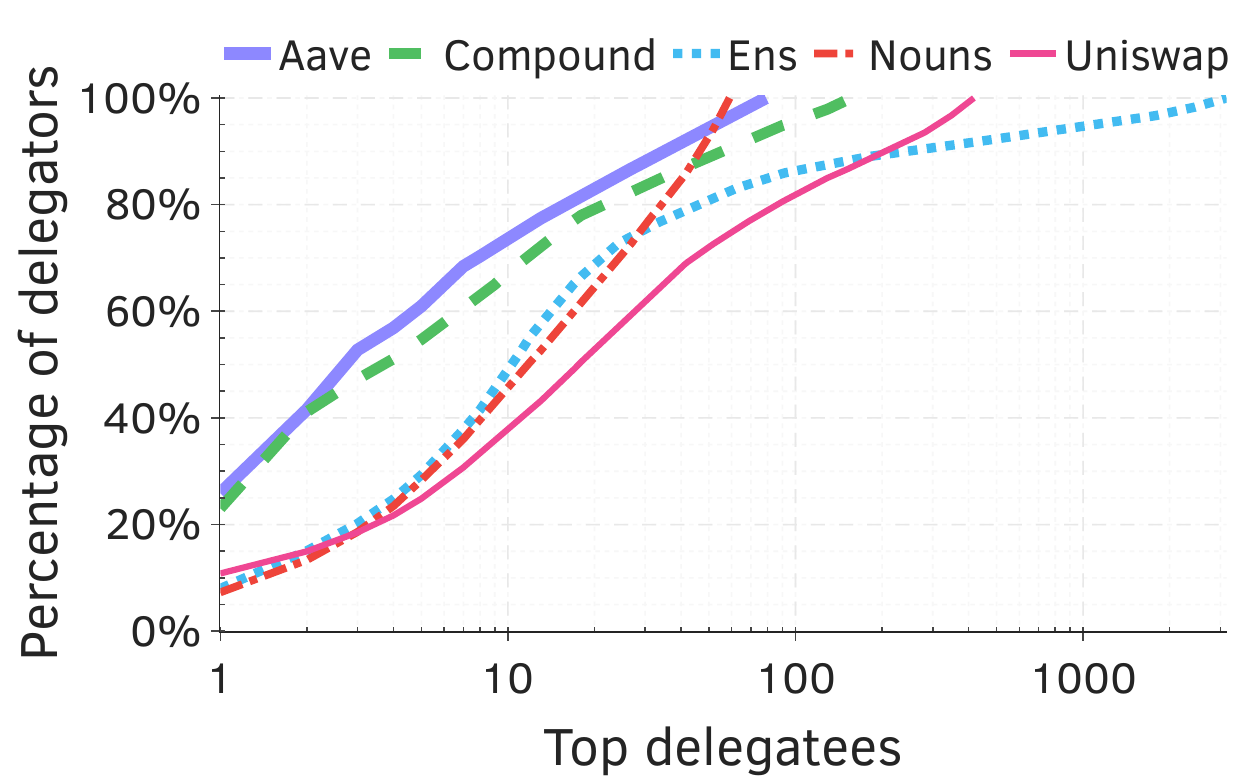}
        \caption{Top delegators ($>$ 1).}
        \label{fig:top-accounts-delegators-2}
    \end{subfigure}
    \caption{Concentration patterns in DAO governance. Figures (a–d) show CDFs of token holders, delegatees, and delegators across major DAOs. Together, these distributions reveal the systemic skew where a few actors dominate holdings and delegated power, while most token holders and delegators remain marginal.}
    \label{fig:top-accounts}
\end{figure*}

We analyze on-chain data from five major governance protocols to assess decentralization in practice. Our study combines two perspectives: (i) concentration of token holdings, voting power, and delegator activity, and (ii) graph-theoretic analysis of delegation networks. 
Together, these reveal how influence accumulates and how delegation structures shape governance outcomes.

\subsection{Token and Power Concentration in DAOs}

Next, we investigate how concentration manifests across multiple layers of DAO governance. 
Figure~\ref{fig:top-accounts-holders} shows that token ownership is heavily skewed in most protocols, with Gini coefficients above $0.99$ for Aave, Compound, ENS, and Uniswap, signaling near-complete inequality in holdings. 
The corresponding Nakamoto coefficients confirm this extreme imbalance: as few as $1$--$8$ addresses can control one-third to one-half of the token supply in ENS and Uniswap, while Compound requires only $19$ addresses to surpass the 50\% threshold. 
By contrast, Nouns displays comparatively broader distribution (Gini $=0.75$), where $9$ and $31$ holders are needed to control 50\% and 66\% of the supply, respectively.

Figure~\ref{fig:top-accounts-delegates} shows that this imbalance is amplified once tokens are delegated and translated into actual voting power. 
Here, the Gini coefficients for voting power all exceed $0.94$, with Compound, ENS, and Uniswap reaching $0.99$, underscoring extreme centralization. 
In ENS, a single delegate controls both one-third and one-half of the delegated power, while Uniswap and Compound require only $16$ and $13$ delegates, respectively, to surpass the 50\% threshold. 
These findings illustrate a clear \emph{rich-get-richer} dynamic: delegation consolidates influence in the hands of a few highly visible actors rather than dispersing it.

Figures~\ref{fig:top-accounts-delegators} and \ref{fig:top-accounts-delegators-2} shift focus to the distribution of \emph{delegators per delegate}, i.e., how many unique accounts have delegated tokens to each delegate. 
When considering all delegations, inequality appears relatively low in Uniswap (Gini $=0.08$) and Compound ($0.16$), where tens of thousands of delegators are spread across many delegates, producing a broad base of support. 
In contrast, ENS again diverges, with a Gini of $0.69$ and a Nakamoto coefficient of only $21$ at the 50\% threshold, meaning that just a few dozen highly visible delegates attract the majority of delegators. 
This highlights that even if many users participate as delegators, their choices are funneled disproportionately toward a small set of popular delegates.

When restricting attention to delegates who receive support from more than one account (Figure~\ref{fig:top-accounts-delegators-2}), concentration rises sharply. 
Uniswap's Gini increases to $0.71$ and ENS exceeds $0.90$, with fewer than $20$ such delegates already attracting two-thirds of all delegators. 
This indicates that diversification of delegations across multiple delegates is rare, and even within this subset, a small elite dominates the attention and support of the community.

Our results reveal that while the presence of many delegators suggests broad participation, the effective distribution of delegator support is heavily concentrated. 
The combination of extreme Gini values and low Nakamoto coefficients across layers confirms systemic vulnerability to capture, where a small number of delegates, backed by a disproportionately large base of accounts, can dictate governance outcomes. 
Such patterns highlight the need for interest-aligned delegation that can redistribute delegator support more equitably for better representative governance in DAOs.

\subsection{Structural Analysis of Delegation Networks}

\subsubsection{Setup}
We model each protocol's delegation structure as a directed graph $G=(V,E)$ where nodes are addresses and a directed edge $u\!\to\!v$ denotes $u$ delegating tokens (i.e., voting power) to $v$. Self-loops ($u\!\to\!u$) represent self-delegation. Unless noted otherwise, connectivity and path-length statistics are computed on the largest undirected connected component (UG LCC). We provide details of our graph network on Table~\ref{tab:delegation-graph-summary} in \S\ref{sec:graph_appendix}.

\subsubsection{Global structure and common patterns}
Across protocols, delegation graphs are \emph{extremely sparse} (densities $10^{-5}\text{--}10^{-4}$) and highly \emph{fragmented} into thousands of weakly connected components (WCCs), with the largest WCC containing at most $6$--$10\%$ of nodes (Aave: $9.75\%$, ENS: $6.69\%$, Uniswap: $1.21\%$, Compound: $2.28\%$, Nouns: $1.71\%$). Degree assortativity is negative in all cases (e.g., ENS: $-0.34$), a hallmark of \emph{disassortative} ``hub-and-spoke'' organization where many low-degree delegators connect to a few high-degree delegates. Clustering and transitivity are near zero, further supporting a star-forest topology. Short diameters ($2$–$6$) and average shortest paths around $2$–$2.7$ within the LCCs are consistent with shallow hub-centric structures. Finally, out-degree distributions are almost degenerate at $1$ (median and $p90$ out-degree $=1$ throughout), reflecting the common practice that a delegator assigns only to a single delegate. This is a constraint of these protocols.

\subsubsection{Protocol-specific insights}
Below we discuss the insights for each of the protocols individually.

\paraib{Aave}
With $6{,}397$ nodes and $6{,}259$ edges (self-loops: $4{,}808$), Aave has the largest WCC share among the five ($9.75\%$), suggesting somewhat tighter connectivity among active delegators than in Compound or Uniswap. The in-degree tail is heavy (max $=357$), with top delegates including \texttt{0x57a$\cdots$922} ($357$), \texttt{0x329$\cdots$ed4} ($216$), and \texttt{kuiqian.eth} ($154$). Degree assortativity is mildly negative ($-0.11$), and the LCC has diameter $6$ and average path length $2.66$, consistent with a constellation of medium hubs rather than a single super-hub.

\paraib{Compound}
Compound's graph ($15{,}230$ nodes, $15{,}052$ edges; self-loops: $12{,}481$) is very fragmented (largest WCC $2.28\%$). A small set of institutional delegates dominates the in-degree: \texttt{a16z} ($344$ and $245$), \texttt{Polychain} ($250$ and $200$), followed by \texttt{Gauntlet} ($118$). Reciprocity is negligible ($0.0001$) and assortativity is negative ($-0.095$). The LCC is shallow (diameter $4$, average path $2.01$), indicating a classic hub-and-spoke driven by branded delegates.

\paraib{ENS}
ENS exhibits the largest scale ($115{,}600$ nodes, $114{,}467$ edges; self-loops: $32{,}783$) and the strongest hub dominance: the top in-degree reaches $6{,}614$, with further hubs at $5{,}840$, $4{,}230$, etc. Disassortativity is pronounced ($-0.34$), and the LCC remains shallow (diameter $6$, average path $2.53$). This is consistent with our CDF-based results: many accounts delegate to a few highly visible ENS figures, creating a particularly concentrated ``attention funnel'' toward prominent delegates.

\paraib{Nouns}
Despite its smaller size ($1{,}112$ nodes; $964$ edges; self-loops: $615$), Nouns maintains the same qualitative structure: sparse, fragmented (largest WCC $1.71\%$), and disassortative ($-0.21$). The in-degree maximum is modest ($18$), and the LCC is extremely shallow (diameter $2$, average path $1.90$). The absence of very large hubs suggests that concentration in Nouns manifests more via token distribution and voting power than via delegator counts per delegate.

\paraib{Uniswap}
Uniswap ($49{,}926$ nodes; $49{,}635$ edges; self-loops: $45{,}664$) has extreme self-delegation and fragmentation (largest WCC $1.21\%$). A few recognizable entities absorb large delegator counts (e.g., \texttt{MultiSig: Univalent} in-degree $454$, \texttt{a16z} $174$, \texttt{Andre~Cronje} $149$, \texttt{Dharma\_HQ} $133$). Assortativity is slightly negative ($-0.03$). The LCC is shallow (diameter $4$, average path $2.37$), again pointing to star-like structures centered on well-known delegates.

\section{Forum Posts Analysis}
\label{sec:off-chain}

We conducted a voting interest analysis based on on-chain traceable governance forum data obtained through voter-entity matching. We filtered all scraped governance forum data using the entity matching list between Tally and the forums, resulting in \num{15259} posts across \num{3981} proposals from 391 unique account addresses. Our analysis consists of two parts. First, to assess the comprehensiveness of the forum data filtered by our matching list, we categorized the proposals to examine the range of topics they cover. Second, we conducted a voters' interest analysis based on this forum data. The following subsections explain the analysis and results for each part.

\subsection{Proposal Categorization}
\paraib{DAO Governance Proposal Taxonomy} Although governance forums cover a wide range of topics, the categorization differs across governance forums and also from third-party services tracking proposals, lacking a unified taxonomy. Thus, we developed our taxonomy for proposal categorization by integrating categories from five DAO governance forums (Aave, Arbitrum, Compound, Ethereum Name Service, Uniswap) and from Messari, a research platform for the cryptocurrency market. Messari has its own governance taxonomy, which classifies all proposals by categories, importance, and sentiment levels. Our taxonomy builds on Messari's taxonomy structure and aims to comprehensively cover the proposal categories defined by each DAO that are not captured by Messari. First, we listed all categories from 14 \glspl{DAO} as well as categories and subcategories from Messari, categorizing similar names into 48 subcategories. When no similar category existed, we assigned a unique subcategory name. We then consolidated these subcategories into 11 categories: \stress{Governance}, \stress{Treasury \& Budgeting}, \stress{Communications, Organization \& Service Providers},  \stress{Protocol Upgrades \& Releases}, \stress{Parameter Change},  \stress{Integrations \& Assets},  \stress{Research \& Development},  \stress{Token Operations}, \stress{Security \& Incident Response}, and \stress{Legal \& Regulatory}. This taxonomy fully covers all categories appearing in each DAO governance forum and Messari.

\begin{figure}[t]
\centering
\includegraphics[width=1\onecolgrid]{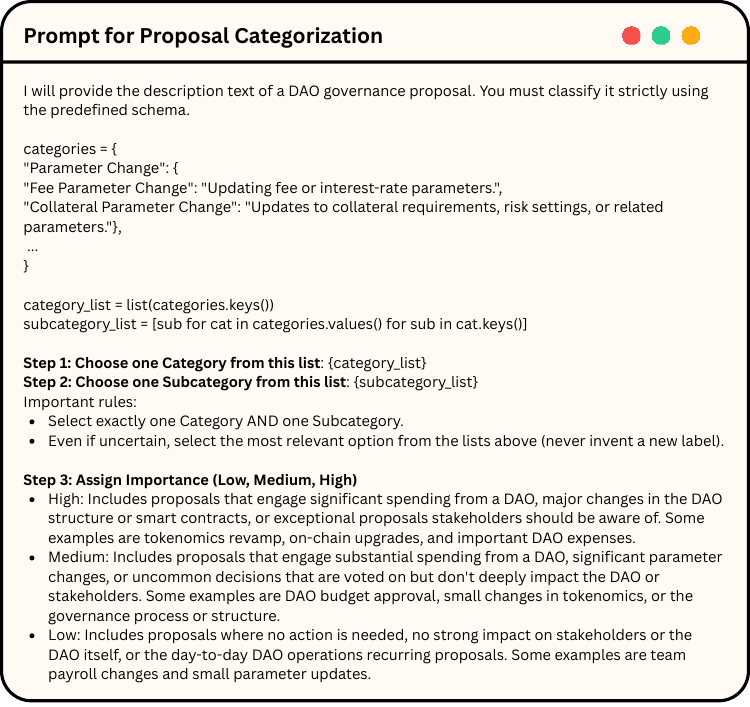}
\caption{Prompt used for Proposal Categorization}
\label{fig:prompt-categorization}
\end{figure}

\paraib{Category and Importance Assignment}
Based on our proposal taxonomy, we assigned each proposal a category, subcategory, and importance level (low, medium, high). For this task, we used the GPT-5 model via the OpenAI API. We followed Messari’s governance taxonomy to define importance and incorporated it into the prompt (see Figure \ref{fig:prompt-categorization}).

\begin{figure}[t]
\centering
\includegraphics[width=1\onecolgrid]{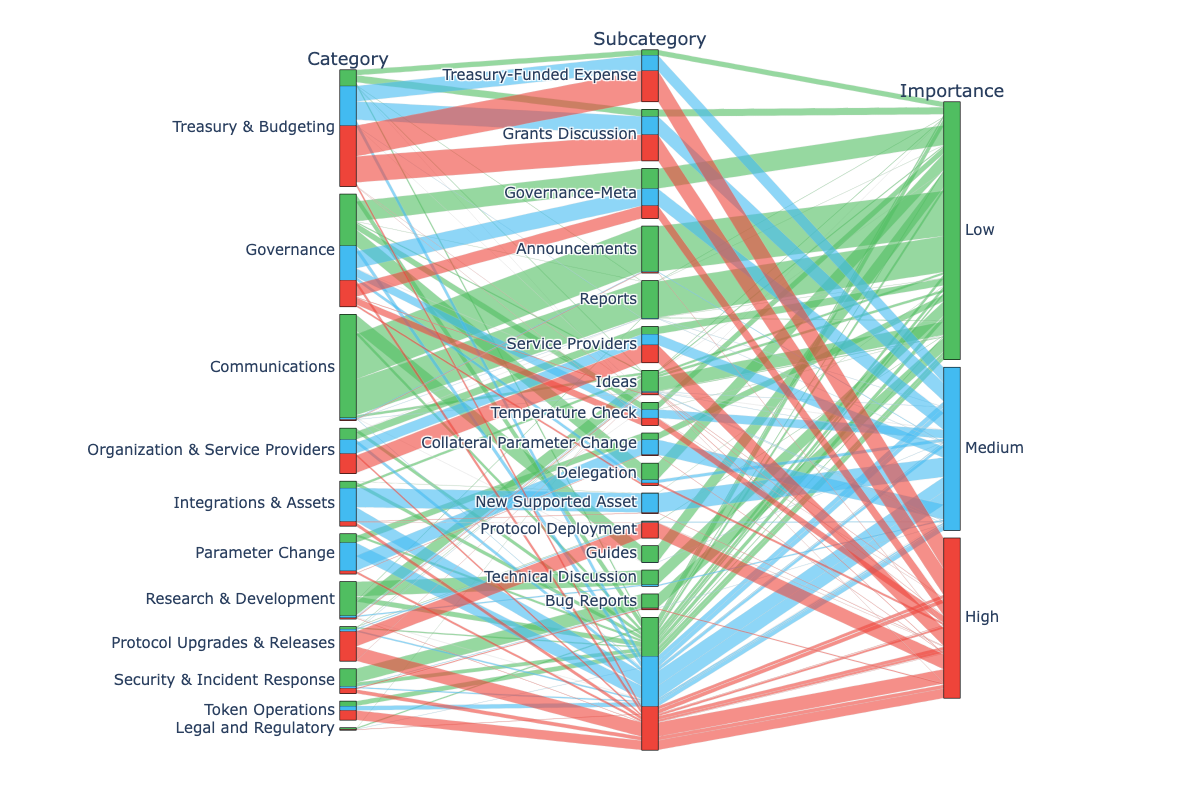}
\caption{Categorization of proposals in our analyzed forum data, using the proposed taxonomy (Category; Subcategory; and Importance)}
\label{fig:sankey-diagram}
\end{figure}

\paraib{Category Distribution}
Figure \ref{fig:sankey-diagram} presents the categorization of the forum data that forms the basis of our subsequent analysis, covering 3981 proposals. Table~\ref{tab:category-summary} further categorizes these matched posts using our proposed taxonomy, showing the distribution of proposal topics and their relative importance across governance discussions.

\begin{table*}[t]
\centering

\caption{Summary of categorized results in our analyzed forum
data.}
\resizebox{\textwidth}{!}{%
\begin{tabular}{lrrrr}
\toprule
\multirow{2}{*}{\thead{Category}} & 
\multirow{2}{*}{\thead{\# of Proposals}} & \multicolumn{3}{c}{\thead{\# of Importance}}\\
&&
\thead{High} & 
\thead{Medium} & 
\thead{Low} \\
\midrule
Treasury \& Budgeting & 798 & 415 (52.0\%) & 273 (34.2\%) & 110 (13.8\%) \\
Governance & 771 & 181 (23.5\%) & 236 (30.6\%) & 354 (45.9\%) \\
Communications & 726 & 5 (0.7\%) & 12 (1.7\%) & 709 (97.7\%) \\
Organization \& Service Providers & 309 & 135 (43.7\%) & 97 (31.4\%) & 77 (24.9\%) \\
Integrations \& Assets & 306 & 30 (9.8\%) & 230 (75.2\%) & 46 (15.0\%) \\
Parameter Change & 274 & 20 (7.3\%) & 194 (70.8\%) & 60 (21.9\%) \\
Research \& Development & 254 & 7 (2.8\%) & 14 (5.5\%) & 233 (91.7\%) \\
Protocol Upgrades \& Releases & 234 & 202 (86.3\%) & 17 (7.3\%) & 15 (6.4\%) \\
Security \& Incident Response & 167 & 34 (20.4\%) & 13 (7.8\%) & 120 (71.9\%) \\
Token Operations & 128 & 65 (50.8\%) & 30 (23.4\%) & 33 (25.8\%) \\
Legal and Regulatory & 14 & 3 (21.4\%) & 1 (7.1\%) & 10 (71.4\%) \\
\bottomrule
\end{tabular}}
\label{tab:category-summary}
\end{table*}

\subsection{Voter Interest Discovery}
\begin{figure}[t]
\centering
\includegraphics[width=1\onecolgrid]{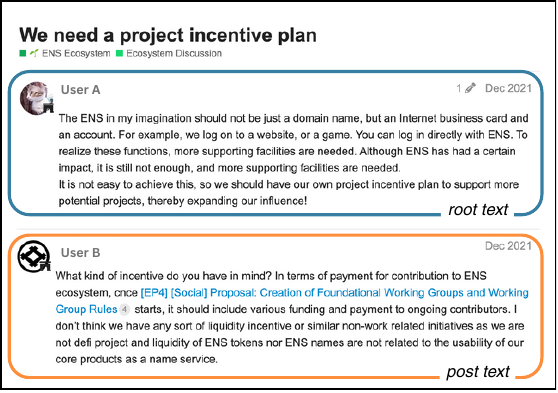}
\caption{Example of a proposal in the governance forum, considering both the voter's post (\textit{post\_text}) and the original context from the proposal’s initial post (\textit{root\_text}).}
\label{fig:proposal-example}
\end{figure}

\paraib{Extracting Interests from Forum Discussions}
To capture voter interests, we began our analysis by extracting keywords from voters’ posts from the forum data. Figure \ref{fig:proposal-example} shows an example of discussion threads within a proposal. Each proposal begins with an initial post that outlines the proposal content, followed by responses from other voters within the thread. Voter responses to proposals reveal the topics they are interested in regarding DAO governance, as well as their positions on each argument. Moreover, the initial post is essential for grasping their interest in comprehending the original context behind their responses. Therefore, we incorporated both the voter post and the original post in the analysis. This approach is aligned with Forum-LDA, which was previously proposed for forum discussion analysis \cite{chen2017forum}. It models root and response posts together to address the sparsity problem of response posts for extracting more coherent forum topics and voter interests.

\begin{figure}[th]
\centering
\includegraphics[width=1\onecolgrid]{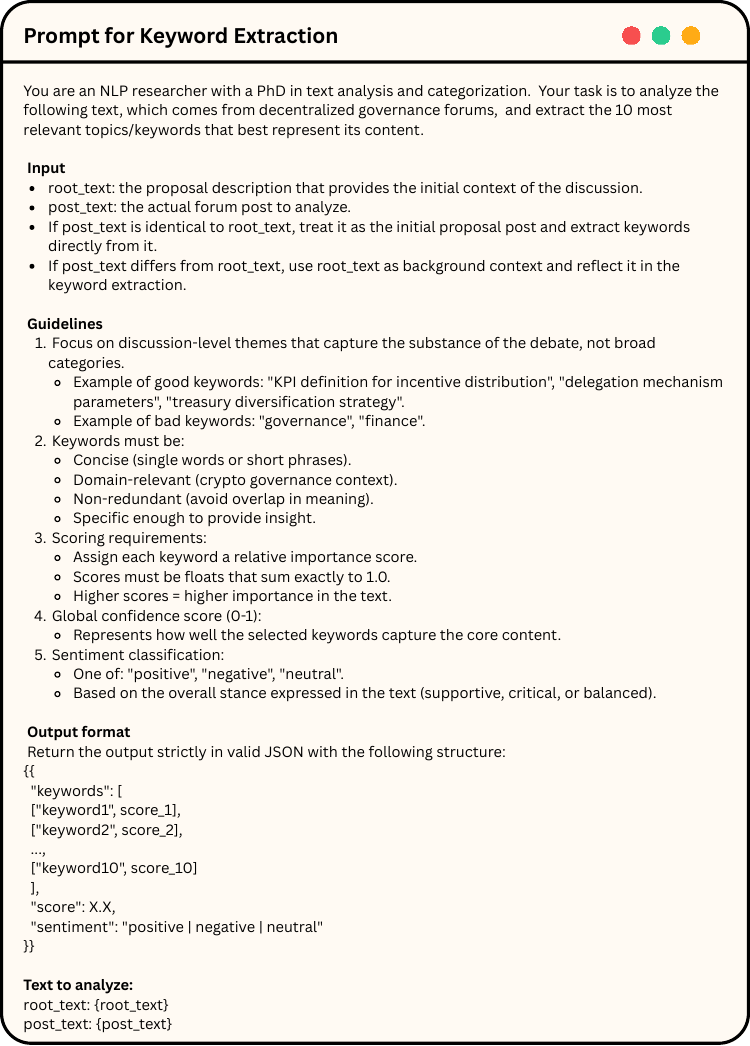}
\caption{Prompt used for Keyword Extraction}
\label{fig:prompt-keywords}
\end{figure}

We employed keyword extraction as a preprocessing step to reduce the dimensionality of forum texts while keeping the interpretable representation for voter interest analysis. Recent studies indicate that methods based on embeddings and \glspl{LLM} surpass traditional statistical or graph-based approaches, offering keywords that are more semantically suitable and varied~\cite{tang2025understanding,Keraghel@DATA}. We therefore compared these two state-of-the-art paradigms. We examined three variants (\textit{gpt-5, gpt-5-mini, and gpt-5-nano}) for \glspl{LLM}, while we used all-MiniLM-L6-v2 for the embedding-based model. For the GPT models, we generated keywords using the prompt shown in Figure \ref{fig:prompt-keywords}, which also guided the models to provide relative importance and confidence scores as indicators for interpretability, along with the stance on the proposal (positive, negative, or neutral). 

As shown in Table \ref{tab:keywords_model}, each model produced keywords at different granularities of information. The \textit{gpt-5} model often produced overly detailed keywords, sometimes capturing specific numeric or temporal details mentioned in the proposal. Conversely, \textit{gpt-5-nano} frequently generated much broader terms, offering greater generality.
The \textit{gpt-5-mini} model was positioned between these extremes, offering keywords that maintained a certain level of detail along with generalization. The embedding-based model \textit{all-MiniLM-L6-v2} produced shorter phrase-like outputs that were similar in results to \textit{gpt-5-nano}, but frequently generated word sequences that lacked semantic coherence. 
Based on these results, we compared clustering results \textit{gpt-5-mini} and \textit{gpt-5-nano} as candidate models and ultimately adopted \textit{gpt-5-nano} considering the interpretability of the resulting clusters.

\begin{table}[t]
\centering

\caption{Top keywords from different models for an example proposal shown in Figure \ref{fig:proposal-example}.}
\label{tab:keywords_model}
\resizebox{\columnwidth}{!}{%
\begin{tabular}{lp{0.75\columnwidth}}
\toprule
\thead{Model} & \thead{Top Keywords}\\
\midrule
\textit{gpt-5} & 
contributor compensation, EP4 Foundational Working Groups, Liquidity incentives rejection, Incentive model clarification, Non-DeFi positioning \\
\midrule
\textit{gpt-5-mini} & 
contributor payment/funding, incentive program for projects, Foundational Working Groups proposal, opposition to liquidity incentives,  ENS as login/account \\
\midrule
\textit{gpt-5-nano} & 
ENS incentive, working groups, working group rules, contributor payments, ecosystem funding \\
\midrule
\textit{all-MiniLM-L6-v2} & 
contribution ens ecosystem, payment contribution ens, liquidity ens tokens, contribution ens, liquidity incentive similar \\
\bottomrule
\end{tabular}}
\end{table}

\begin{figure}[th]
\centering
\includegraphics[width=1\onecolgrid]{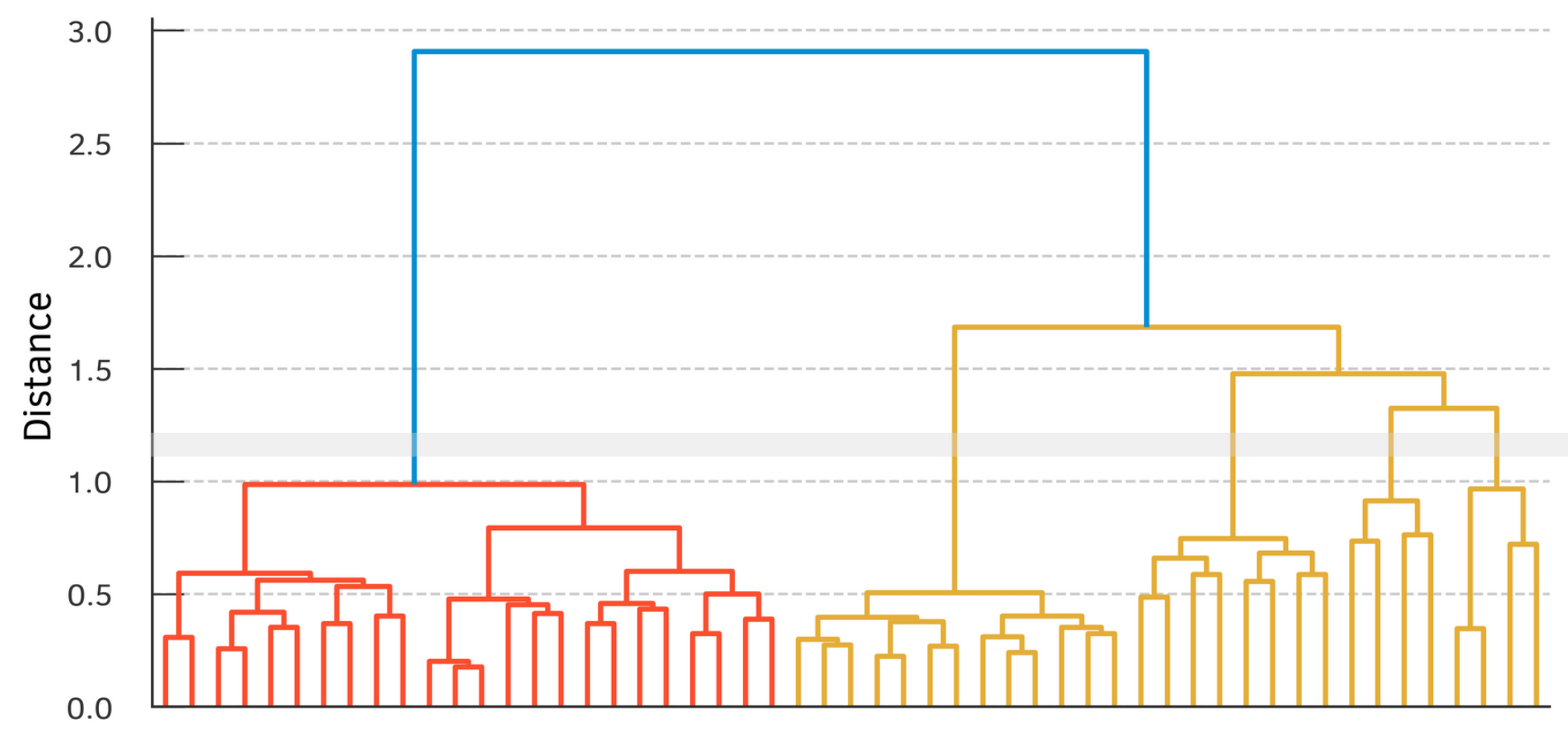}
\caption{Dendrogram from hierarchical clustering}
\label{fig:dendrogram}
\end{figure}

\paraib{Identifying Interest Similarities through Cluster Analysis}
To examine similarities in voters’ interests, we first aggregated the extracted keywords from individual posts to the voter level, creating a representative set of keywords for each voter. After this aggregation step, we converted the keyword sets into semantic vector representations. Traditional keyword frequency methods, such as Bag-of-Words or TF-IDF, treat surface-level variants (e.g., ``Snapshot vote'' vs. “Snapshot voting”) as independent features. In contrast, embedding models map semantically related expressions close to each other in a dense vector space, enabling a more conceptually accurate representation of interests. Specifically, we employed the \textit{all-MiniLM-L12-v2} model, which encodes sentences and short phrases into a 384-dimensional dense vector space suitable for clustering and semantic similarity tasks. This procedure yielded one vector per voter, capturing the aggregated semantics of their expressed interests. We then applied hierarchical clustering to classify voters based on these vectors. We employed Ward’s method as it minimizes within-cluster variance, and the resulting dendrogram is shown in Figure \ref{fig:dendrogram}. We determined the optimal threshold where distances between clusters are greater than 1, resulting in five distinct clusters.

\begin{figure}[th]
\centering
\includegraphics[width=1\onecolgrid]{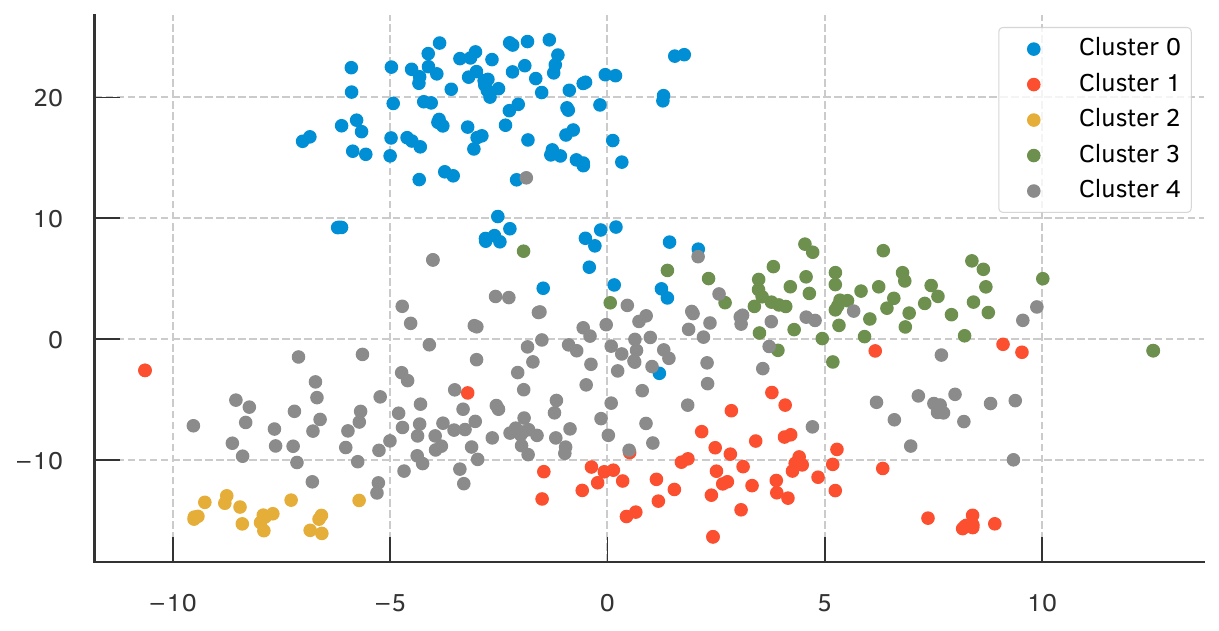}
\caption{t-SNE plot visualizing cluster assignments of voters.}
\label{fig:t-SNE}
\end{figure}

To inspect the similarities of these clusters, we projected the high-dimensional voter interest vectors into a two-dimensional space using t-SNE, as illustrated in Figure \ref{fig:t-SNE}. Cluster 0 was the largest group with 156 voters, followed by Cluster 3 with 111 voters and Cluster 2 with 78 voters, while Cluster 1 was the smallest with 46 voters.

\begin{figure}[th]
\centering
\includegraphics[width=1\onecolgrid]{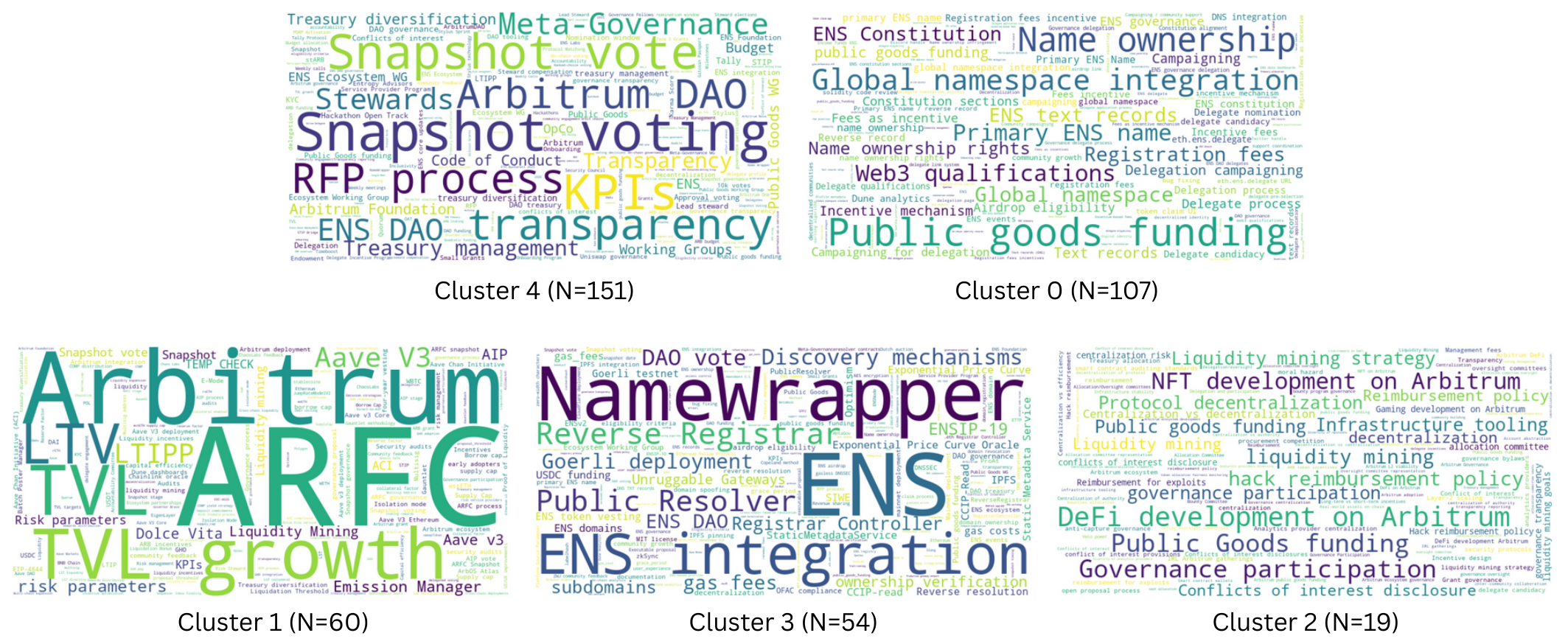}
\caption{Word clouds of top keywords characterizing each voter cluster.}
\label{fig:word-clouds}
\end{figure}

Finally, to characterize each cluster, we generated word clouds of the most representative keywords, presented in Figure \ref{fig:word-clouds}. Cluster 4 corresponds to \stress{Finance-Driven} voters, with a strong focus on treasury management, accountability, and performance metrics. Cluster 0 can be seen as \stress{Governance-Driven} voters, highlighting concerns around governance principles, name ownership, and the allocation of public goods funding. Cluster 1 captures \stress{Arbitrum DeFi-Driven} voters, whose discussions emphasize DeFi parameters, liquidity incentives, and risk management. Cluster 3 reflects \stress{ENS Ecosystem-Driven} voters, oriented toward ENS operations, technical upgrades, and working group governance. Lastly, Cluster 2 illustrates \stress{Innovation-Driven} voters, characterized by interest in emerging initiatives such as NFTs and broader decentralization theme.

\section{Concluding Discussion}
\label{sec:conclusion}

We presented a large-scale empirical study of delegation in \gls{DAO} governance based on off-chain discussions from 14 DAO forums. Our findings show that delegation, while intended to increase participation, often amplifies existing power imbalances. These findings challenge the assumption that delegation, as currently implemented, naturally leads to more representative governance.
Our study offers several implications. For researchers, it demonstrates the importance of examining delegation not only as a structural mechanism but also through the lens of interest alignment between communities and their representatives. By analyzing the conversational and ideological contexts that shape governance preferences, we show that understanding delegation requires going beyond purely quantitative or on-chain perspectives. For practitioners, our results highlight the risks embedded in current ranking-based delegate discovery interfaces, which reinforce visibility biases and entrench existing elites. Designing more transparent, informative, and alignment-aware mechanisms could better guide token holders toward representatives who reflect their priorities, thereby improving participation, reducing voter apathy, and enhancing the overall fairness and representativeness of \gls{DAO} governance.



\bibliographystyle{splncs04}
\bibliography{references}

\appendix
\renewcommand{\thesection}{Appendix \arabic{section}}

\section{Delegation Graph Statistics}
\label{sec:graph_appendix}

Table~\ref{tab:delegation-graph-summary} reports detailed structural statistics for delegation graphs across the five studied protocols. 
Each graph models addresses as nodes and delegation relationships as directed edges, with self-loops capturing self-delegations. 
We compute standard network metrics to highlight structural properties.

\begin{table*}[t]
\centering
\caption{Delegation graph summary per protocol. Nodes are addresses; directed edges encode delegations; self-loops are self-delegations. All connectivity and path-length metrics are computed on the largest undirected connected component (LCC).}
\resizebox{\textwidth}{!}{%
\begin{tabular}{lrrrrrrrrr}
\toprule
\thead{Protocol} & \thead{Nodes} & \thead{Edges} & \thead{Self-loops} & \thead{Largest WCC} & \thead{Max In-deg.} & \thead{Reciprocity} & \thead{Assortativity} & \thead{Avg Path (LCC)} & \thead{Diameter (LCC)} \\
\midrule
Aave     & 6{,}397  & 6{,}259  & 4{,}808 & 9.75\% & 357 & 0.0006 & $-0.1103$ & 2.662 & 6 \\
Compound & 15{,}230 & 15{,}052 & 12{,}481 & 2.28\% & 344 & 0.0001 & $-0.0948$ & 2.011 & 4 \\
ENS      & 115{,}600& 114{,}467& 32{,}783 & 6.69\% & 6{,}614 & 0.0018 & $-0.3395$ & 2.530 & 6 \\
Nouns    & 1{,}112  & 964     & 615   & 1.71\% & 18  & 0.0000 & $-0.2120$ & 1.895 & 2 \\
Uniswap  & 49{,}926 & 49{,}635 & 45{,}664 & 1.21\% & 454 & 0.0000 & $-0.0335$ & 2.372 & 4 \\
\bottomrule
\end{tabular}}
\label{tab:delegation-graph-summary}
\end{table*}

\section{DAO Governance Proposal Taxonomy} 
\label{sec:taxonomy}

To systematically analyze governance posts across \glspl{DAO}, we construct a taxonomy that unifies heterogeneous forum categories into a consistent schema. 
Tables~\ref{tab:taxonomy_1} and~\ref{tab:taxonomy_2} present this taxonomy, covering governance processes, treasury and budgeting, protocol upgrades, integrations, security, organizational aspects, token operations, research, and communications.

\begin{table*}[t]
\centering
\caption{DAO Governance Proposal Taxonomy Part 1.}
\setlength{\tabcolsep}{3pt}
\resizebox{\textwidth}{!}{%
\begin{tabular}{llp{4.5cm}p{4.5cm}}
\toprule
\thead{Category} & \thead{Subcategory} & \thead{Original Category/Subcategory} & \thead{Definition} \\
\midrule
Governance & Governance-Meta & Governance (Aave, Arbitrum), Governance-Meta (Uniswap, ENS), Site Feedback (Uniswap, Aave), DAO-Wide (Uniswap), Governance Process (Compound) & Meta discussion about how governance/forum operates. \\
\midrule
Governance & Proposals & Proposals (Compound, Wormhole, Arbitrum), Consensus Check (Uniswap), Request for Comment (Uniswap)& Formal or draft proposals submitted for community review and voting.\\
\midrule 
Governance & Delegation & Delegation Pitch (Uniswap), Delegate Platforms (Aave, Wormholes) & Processes for delegates to present themselves and request delegated voting power.\\
\midrule 
Governance & Governance Update & Governance Update (Messari) & Changes to rules, quorums, or frameworks.\\
\midrule 
Governance & Temperature Check & Temperature Check (Uniswap) & Initial community vote to gauge support before a formal proposal.\\
\midrule 
Legal and Regulatory & Regulatory Status & Regulatory Status (Messari) & Disclosures/filings on regulatory posture.\\
\midrule 
Legal and Regulatory & Legal Action & Legal Action (Messari) & Initiating/responding to legal proceedings. \\
\midrule
Parameter Change & Fee Parameter Change & Fee Parameter Change (Messari) & Updating fee or interest-rate parameters.\\
\midrule
Parameter Change & Collateral Parameter Change &
Collateral Parameter Change (Messari) & Updates to collateral requirements, risk settings, or related parameters.\\
\midrule
Treasury \& Budgeting & Grants Discussion & Aave Grants DAO (Aave), Grants Discussion (Arbitrum), Grants (Compound), DAO Programs \& Initiatives (Arbitrum) & Creation or management of grants and funding programs for contributors.\\
\midrule
Treasury \& Budgeting & Treasury-Funded Expense & Treasury-Funded Expense (Messari) & Requests for funding specific projects or operational costs from the treasury.\\
\midrule
Treasury \& Budgeting & Debt Funding & Debt Funding (Messari) & Proposals to raise funds through debt or borrowing mechanisms.\\
\midrule
Treasury \& Budgeting & Public Goods & Public Goods (ENS) & Dedicated funding for public goods.\\
\midrule
Protocol Upgrades \& Releases & Core Client Release & Core Client Release (Messari) & Release of reference clients or core software.\\
\midrule
Protocol Upgrades \& Releases & Fork & Fork (Messari) & Creating a protocol/network fork/split.\\
\midrule
Protocol Upgrades \& Releases & On-chain Upgrade & On-chain Upgrade (Messari) & Protocol logic change executed on-chain. \\
\midrule
Protocol Upgrades \& Releases & Protocol Deployment & Contract Deployment (Messari) & Deployment of new smart contracts or protocol modules.\\
\midrule
Protocol Upgrades \& Releases & Branding & New Brand or Rebrand (Messari) & Brand changes or rebrands.\\
\midrule
Protocol Upgrades \& Releases & Launch & Mainnet Launch (Messari), v4 Launch (Uniswap) & Launch of the mainnet or major new protocol version.\\
\midrule
Protocol Upgrades \& Releases & Other Change & Other Project Change (Messari), Other Network Change (Messari) &Changes that don’t fit into standard protocol upgrade categories.\\
\midrule
Protocol Upgrades \& Releases & Ideas & Ideas (Compound) &  Informal brainstorming or idea-sharing before formal proposals. \\
\bottomrule
\end{tabular}
}
\label{tab:taxonomy_1}
\end{table*}

\begin{table*}[t]
\centering
\caption{DAO Governance Proposal Taxonomy Part 2.}
\setlength{\tabcolsep}{3pt}
\resizebox{\textwidth}{!}{%
\begin{tabular}{llp{4.5cm}p{4.5cm}}
\toprule
\thead{Category} & \thead{Subcategory} & \thead{Original Category/Subcategory} & \thead{Definition} \\
\midrule
Integrations \& Assets & New Collateral Asset & New Collateral Asset (Messari)& Authorize an asset as eligible collateral.\\
\midrule
Integrations \& Assets & New Supported Asset & New Supported Asset (Messari) & Add a new asset to a market/listing.\\
\midrule
Integrations \& Assets & New Markets & New Markets (Compound), Conditional Funding Markets (Uniswap) & Discussion of new assets \& markets for the protocol. \\
\midrule
Integrations \& Assets & Wallet Integration & Wallet Integration (Messari) & Integrate external wallets or SDKs.\\
\midrule
Integrations \& Assets & Oracle Integration & Oracle Integration (Messari) & Integrate oracle solutions.\\
\midrule
Integrations \& Assets & Scaling Solutions Integration & Scaling Solutions Integration (Messari) & Integrate scaling solutions. \\
\midrule
Integrations \& Assets & Cross-Chain Bridge Integration & Cross-Chain Bridge Integration (Messari) & Integrate a cross-chain bridge to transfer assets across different blockchains.\\
\midrule
Security \& Incident Response & Risk & Risk (Aave) & Topics that are risk-related. \\
\midrule
Security \& Incident Response & Bug Reports & Bug Reports (ENS) & Community-reported bugs (e.g., ENS Manager).\\
\midrule
Security \& Incident Response & Bug Bounty Program & Bug Bounty Program (Messari) & Create/modify incentives for vulnerability reporting.\\
\midrule
Security \& Incident Response & Bug Disclosure & Bug Disclosure (Messari) & Formal disclosures of bugs or vulnerabilities found in the system.\\
\midrule
Organization \& Service Providers & Project Team & Project Team (Messari), Security Council Elections (Arbitrum) & Team composition, hiring or working groups, elections.\\
\midrule
Organization \& Service Providers & Service Providers & Service Provider Program (ENS), Service Providers (Uniswap) & Service-provider onboarding and admin.\\
\midrule
Organization \& Service Providers & Partnership & Partnership (Messari) & Formal partnerships to expand ecosystem reach.\\
\midrule
Token Operations & Distribution & Airdrop (Messari), Liquidity Mining Program (Messari) & Token distributions to users/liquidity providers.\\
\midrule
Token Operations & New Token & New Token (Messari) & Create or designate a new governance/utility token. \\
\midrule
Token Operations & Token Sale & Private Token Sale (Messari), Public Token Sale (Messari) & Primary sales to investors or the public.\\
\midrule
Token Operations & Token Swap & Token Swap (Messari) & Swapping DAO-held tokens with counterparties.\\
\midrule
Token Operations & Other Token Change & Other Token Change (Messari) & Any token-level change not covered above.\\
\midrule
Token Operations & Listings & Centralized Exchange Listing (Messari) & Listing tokens. \\
\midrule
Token Operations & Supply Actions & Supply Unlock (Messari), Non-programmatic Supply Burn (Messari), Non-programmatic Supply Mint (Messari) & Non-programmatic mint/burn or scheduled unlocks.\\
\midrule
Research \& Development & Development & Protocol Development (Compound), ENS Development (ENS), Development (Aave) & Technical development for protocol components.\\
\midrule
Research \& Development & Research \& Development Updates & Research \& Development Collective (Arbitrum) & Updates and outputs from R\&D working groups.\\
\midrule
Research \& Development & Technical Discussion & Technical Discussion (Arbitrum) & Forum space for technical conversations or debates.\\
\midrule
Communications & Announcements & Announcements (Arbitrum, Wormhole) & Official announcements and key updates shared with the community.\\
\midrule
Communications & Guides & Guides (Compound), Learning Center (Aave) & Educational guides and learning resources.\\
\midrule
Communications & Reports & Report (Messari) & Periodic reports summarizing DAO activities and outcomes.\\
\bottomrule
\end{tabular}
}
\label{tab:taxonomy_2}
\end{table*}

\end{document}